\documentstyle[sprocl_mod,epsfig,11pt]{article}

\def\be{\begin{equation}}
\def\ee{\end{equation}}
\def\bea{\begin{eqnarray}}
\def\eea{\end{eqnarray}}
\newcommand{\bq}{\begin{equation}} 
\newcommand{\eq}{\end{equation}} 
\newcommand{\ba}{\begin{eqnarray}} 
\newcommand{\ea}{\end{eqnarray}}

\newcommand{\mz}{M_{_Z}}
\newcommand{\afba}[1]{A^{#1}_{_{\rm FB}}}

\def\be{\begin{equation}}
\def\ee{\end{equation}}
\def\ba{\begin{eqnarray}}
\def\ea{\end{eqnarray}}

\begin{document}

\renewcommand{\thefootnote}{\fnsymbol{footnote}}
\setcounter{footnote}{0}

\thispagestyle{empty}
\vspace*{-2cm}
\begin{flushright}
{
LC-TH-2000-008
\\
February 2000
\\
hep-ph/0002054
}
\end{flushright}

\bigskip

\title{RADIATIVE CORRECTIONS TO $e^+e^-\to \bar{f} f$
\footnote{{\it Talk presented by M.~Jack, contribution to the Proceedings of the 2nd Joint ECFA/DESY Study on Physics and Detectors for a Linear Electron-Positron Collider, held at Orsay, Lund, Frascati, Oxford, and Obernai from April 1998 until Oct 1999, to appear as DESY report 123F, ``Physics Studies for a Future Linear Collider''
(R.~Heuer, F.~Richard, P.~Zerwas, eds.)}
}}
\author{P. CHRISTOVA\footnote{Supported by Bulgarian Foundation for
Scientific Research with grant $\Phi$--620/1996.}
}
\address{
Lab. for Nuclear Problems, Joint Institute for Nuclear Research, 
\\
Dubna, Russia
\\ 
E-mail: penchris@nusun.jinr.ru}
\author{M. JACK, S. RIEMANN, and T. RIEMANN}
\address{DESY Zeuthen, Platanenallee 6,
\\ 
D-15738 Zeuthen, Germany
\\ 
E-mails: jack@ifh.de, riemanns@ifh.de, riemann@ifh.de}

\maketitle
\abstracts{
The past ten years of physics with $e^+e^-$ colliding 
experiments at LEP and SLAC \cite{Sirlin:1999lp} have shown the
success of these experiments on not only impressively proving
the theoretical predictions of the Standard Model ({\tt SM}),
but also to help provide stringent bounds on physics beyond 
the {\tt SM}. With this experience in mind, there appear two 
equally fascinating opportunities for studying fermion-pair 
production processes at a future Linear Collider (LC). 
On the one hand,   
performing high precision measurements to the {\tt SM},
for example, when running with high luminosity at the 
$Z$ boson resonance, could be a quick and feasible
enterprise in order to pin down the symmetry breaking 
mechanism of the electroweak sector through indirectly 
determining the masses of a light {\tt SM} or {\tt MSSM} 
Higgs boson or supersymmetric particles via
virtual corrections. 
On the other hand, looking for such particles in direct 
production or other `New Physics' effects at energies 
between, for example, roughly 500 and 800 GeV will naturally 
be the main motivation to pursue the challenging endeavor of 
building and utilizing such a unique facility. 
These two scenarios for the LC shall be sketched here, with 
particular emphasis on the semi-analytical program {\tt ZFITTER} 
for fermion-pair production in comparison with numerical programs 
like {\tt TOPAZ0}, {\tt KK2f}, and others.
}

\setcounter{footnote}{0}
\renewcommand{\thefootnote}{\alph{footnote}}

%                         Text
%==========================================================================
\section{\bf Introduction
\label{sec_intro}
}
%--------------------------------------------------------------------------
%
A future Linear Collider (LC) running with high luminosities 
at energies up to $500\ldots 800\,\mbox{GeV}$ will demand a 
dedicated effort not only on its experimental realization, 
but also from the theoretical side on predicting observables 
under experimentally realistic conditions with an unprecedented 
precision \cite{Boudjema:1996qg,Accomando:1997wt}. For this, 
theory has applied quantum field theory successfully to 
calculate quantum corrections in a perturbative approach, in 
order to accurately predict or confirm high energy observables 
in the past, which will be even more demanding for the special 
case of the LC with its high resolution power. 
The proof \cite{NobelPrize:1999} that this can be done on the solid ground of 
gauge theories as renormalizable, unitarity-conserving 
quantum field theories for all three microscopically 
observed forces in nature -- the electromagnetic, 
weak, and strong interaction -- was rewarded just recently, 
underlining the validity and practical 
applicability of our modern-day theoretical particle physics 
description. In practice, this also resulted in such pioneering work 
-- just to show some examples -- as giving limits on mass differences 
in the electroweak (EW) sector of the {\tt SM}, 
resulting in indirect top mass bounds from LEP, 
calculating radiative corrections to the weak vector boson masses, 
shown at LEP, or providing first computational tools indispensable for 
today's practically involved, Feynman-diagrammatic calculations 
using computer algebra techniques \cite{NobelPrize2:1999}.

The goal of this paper in the context of an $e^+e^-$ LC now is 
to briefly outline the, in this respect still 
interesting and fruitful physics potential 
of the `classical' fermion-pair production processes, 
$e^+e^-\to \bar{f}f$, at these energies and luminosities 
\cite{Accomando:1997wt}, but also the theoretical 
implications and resulting necessities stressed for an update of 
existing numerical programs, especially for 
the case of the semi-analytical code {\tt ZFITTER} 
\cite{Bardin:1989all,Christova:1999cc,DESY99070,Christova:1998tc,DESY99037,SITGES99,Jack:1999qh,Jack:1999ox}.

It is absolutely clear that the focus at such a machine will be 
on physics at these high energies, for which it is envisaged, 
but also a quick, high-luminosity run at the $Z$ boson resonance 
could be an interesting extra option (Giga-Z), thus substantially 
increasing the experimental precision on $Z$ lineshape observables 
and the {\tt SM} parameters, on which they are sensitive, with a
manageable extra effort
\cite{Moenig:1999ox,EWsummary:1999ox,Heinemeyer:1999ox,Weiglein:1999si}.
Both situations --  the Giga-Z option and the high-energy run -- 
shall be discussed now for the fermion-pair production case,
with special emphasis on what this means for the program 
{\tt ZFITTER} \cite{Bardin:1992jc,DESY99070} 
in comparison with other available 2-fermion codes
\cite{Beenakker:1991mb,Burgers:BHM,Montagna:1998kp,Jadach:1994yvx,Jadach:1999kkkz}.

%==========================================================================
\subsection{\bf High precision measurements to the SM and MSSM
\label{zbb}
}
%--------------------------------------------------------------------------
%
First, the Giga-Z option:
In \cite{Moenig:1999ox} it was demonstrated that with a factor of 100 or so 
higher statistics than at LEP running on the $Z$ boson resonance
\footnote{In comparison to SLD at SLAC, it would be even a factor
of roughly 2000.
}
-- corresponding to a luminosity of 
${\cal L}\sim 5\cdot 10^{33}cm^{-2}s^{-1}$ or roughly $10^9$ 
hadronic $Z$ boson decays after just a few months of running -- 
especially fermion-pair production asymmetries like $A_{LR}$ 
or the polarized $b\bar{b}$ forward-backward asymmetry 
$A^b_{FB}$ could be measured with very high precision
when using one or both beams polarized. This latter condition 
together with good $b$-tagging techniques and 
the collected experiences at LEP and SLD 
should help to keep the systematic errors 
under control.
The implication of this from the theoretical side 
on extracted {\tt SM} parameters, like e.g. on the $W$ boson
mass, $M_W$, or the effective weak mixing angle, 
$\sin^2\theta_{eff}$, was nicely illustrated in \cite{Heinemeyer:1999ox},
with expected experimental accuracies of $\Delta M_W = 6\,\mbox{MeV}$ 
or $\Delta\sin^2\theta_{eff} = 4\times 10^{-5}$ at the 
Giga-Z. This is to be compared with the presently 
achievable total experimental errors by the end of LEP \cite{Moenig:1999ox} 
of $\Delta M_W = 40\,\mbox{MeV}$ or 
$\Delta\sin^2\theta_{eff} = 1.8\times 10^{-4}$.
Due to loop corrections, $M_W$ and $\sin^2\theta_{eff}$  
are sensitive to the mass of a light Higgs boson, $M_h$, the 
top quark mass, $m_t$, and in the supersymmetric ({\tt susy}) case, 
also on the {\tt susy} mass scale, $M_{susy}$. These much 
improved experimental values 
for $M_W$ and $\sin^2\theta_{eff}$ thus allow at the Giga-Z 
together with the precise knowledge of $m_t$ an indirect determination
of the mass of a light Higgs boson in the {\tt SM} at the 10\% 
level or strong consistency checks on the 
{\tt SM/MSSM} values of $M_W$ and $\sin^2\theta_{eff}$ 
with $m_t$ and {\tt susy} masses as input parameters \cite{Heinemeyer:1999ox}.

%==========================================================================
\subsection{\bf Virtual Corrections and New Physics Phenomena
\label{newphysics}
}
%--------------------------------------------------------------------------
%
Probably one of the most fascinating applications of fermion-pair production 
processes at higher energies is then the search for 
`New Physics Phenomena' ({\tt NPP}), i.e. observed effects not described 
by the {\tt SM} \cite{newphys:1999sum}. 
This is of course quite actively pursued already at existing 
$e^+e^-$ facilities, giving quite stringent bounds on masses and couplings of 
exchanged `exotic' particles or minimal interaction scales of {\tt NPP}.
With a future LC, however, reaching much higher energies at or nearly at the 
$\mbox{TeV}$ scale and using high luminosities,  
there is the justified hope of touching this `beyond the {\tt SM}' 
domain of particle physics and really uncovering {\tt NPP}.
Examples of such investigations 
\cite{ZWprcontsusy:1999ox,Leptoq:1999lc,Extradim:1999sg} 
are e.g. setting lower limits on 
four-fermion contact interaction scales 
or on masses and couplings of extra heavy neutral or 
charged gauge bosons, $Z'$ and $W'$, 
of {\tt susy} particles in ${\cal R}$ parity violating 
supersymmetric models, or for interaction-unifying models ({\tt GUTs}), 
searches for excited leptons, leptoquarks, preons etc., 
or looking at effects of spin-2 boson exchanges on e.g.
angular cross section distributions in string-inspired,
low-scale quantum gravity models. With a LC, the so far
checked energy region for {\tt NPP} from LEP/SLD can be
extended from typically $O(\mbox{few TeV})$ up to several  
tenths of $\mbox{TeV}$ at a LC. For a summary of these 
activities also refer to \cite{newphys:1999sum,Leptoq:1999lc}.

Another quite interesting application in the context of the 
Giga-Z option could be looking for lepton flavor
number violating $Z$ decays like $Z\to\mu\tau$, $e\tau$, or $e\mu$ when
heavy neutrinos are exchanged in virtual corrections (Dirac or Majorana type).
The estimated branching ratios in the case of $Z\to e\tau$ or $\mu\tau$
could be large enough in some models to be observable at the Giga-Z. 
There is a vast literature on this topic, and also some preliminary 
studies for the LC were presented at this workshop \cite{Zmutau:1999ox}.

%==========================================================================
\section{\bf ZFITTER and other programs at the $Z$ boson resonance
\label{photcorr_Z}   
}
%--------------------------------------------------------------------------
%
The semi-analytical program {\tt ZFITTER} 
\cite{Bardin:1989all,Christova:1999cc,DESY99070} calculates 
observables for fermion-pair production like total cross sections 
and asymmetries and contains analytical formulae with the exact 
one-loop radiative corrections and important higher order effects
for {\tt SM} applications, or alternatively, allowing a model-independent 
approach e.g.~for `New Physics' searches (see Section~\ref{newphysics}).
In the program, the analytical formulae only need to be numerically 
integrated over the final state invariant mass squared, 
$m^2_{f\bar{f}}$,
making the code numerically fast and stable
with the inclusion of a limited number of 
experimentally relevant, kinematical cuts to the final state.
After a recent update of the {\tt ZFITTER} code 
for the special case of combined cuts on 
energies, acollinearity, and acceptance angle 
of leptonic final states \cite{Christova:1999cc,DESY99070} and comparisons 
with other numerical programs 
\cite{Beenakker:1991mb,Montagna:1998kp,Jadach:1999kkkz}, 
the situation
for LEP 1 energies can be stated as quite satisfactory 
\cite{Christova:1998tc,Bardin:1999gt,DESY99037,SITGES99,Jack:1999qh,Jack:1999ox}: 
total cross sections and forward-backward asymmetries 
are now treated better than at the per mil level 
for both cut options -- the invariant mass and the acollinearity cut --
around the $Z$ boson resonance, and even better than $10^{-4}$
at the $Z$ peak itself, illustrated in Table~\ref{tab-acol10-th40} below
(from Table 7 in \cite{DESY99037} and Table 2 in \cite{Jack:1999qh,Jack:1999ox}).
%
%xxxxxxxxxxxxxxxxxxxxxxxxxxxxxxxxxxxxxxxxxxxxxxxxxxxxxxxxxxxxxxxxx 
\begin{table}[htb]
\begin{center}
\renewcommand{\arraystretch}{1.1}
\begin{tabular}{|c||c|c|c|c|c|}
\hline
\multicolumn{6}{|c|}{
{$\sigma_{\mu}\,$[nb] with $\theta_{\rm acol}<10^\circ$}}
\\ 
\hline
$\theta_{\rm acc} = 0^\circ$& $\mz - 3$ & $\mz - 1.8$ & $\mz$ & $\mz + 1.8$ &
$\mz + 3$  \\ 
\hline\hline
  & 0.21932  & 0.46287  & 1.44795  & 0.67725  & 0.39366 \\
{{\tt TOPAZ0}}  & 0.21776  & 0.46083  & 1.44785  & 0.67894  & 0.39491 \\
  & 
{\bf --7.16}    & 
{\bf --4.43}     &
{\bf --0.07}     &
{\bf +2.49}     &
{\bf +3.17}    
\\ 
\hline
  & 0.21928  & 0.46284  & 1.44780  & 0.67721  & 0.39360 \\
{{\tt ZFITTER}}  & 0.21772  & 0.46082  & 1.44776  & 0.67898  &
0.39489 \\ 
  &
{\bf --7.16}     &{\bf --4.40}     &{\bf --0.03 }    &{\bf +2.60}
&{\bf +3.27}   
\\ 
\hline 
\hline
\multicolumn{6}{|c|}{$\afba{\mu}$ with $\theta_{\rm acol}<10^\circ$} \\
\hline
$\theta_{\rm acc} = 0^\circ $& $\mz - 3$ & $\mz - 1.8$ &
$\mz$ & $\mz + 1.8$ & $\mz + 3$  \\ 
\hline\hline
  & --0.28450 & --0.16914  & 0.00033  & 0.11512  & 0.16107 \\
{{\tt TOPAZ0}}  & --0.28158 & --0.16665  & 0.00088  & 0.11385  &
0.15936 \\ 
  & 
{\bf +2.92}    & {\bf +2.49}     &
{{\bf +0.55}}     &{\bf --1.27}
&
{{\bf --1.71}}    \\ 
\hline
  & --0.28497 & --0.16936  & 0.00024  & 0.11496  & 0.16083 \\
{{\tt ZFITTER}}  & --0.28222 & --0.16710  & 0.00083  & 0.11392  & 0.15926 \\
  & {\bf +2.75}    & {\bf +2.27}     & {\bf +0.60}    &{\bf --1.03}
&{\bf --1.56}
%----------------------------
\\
\hline 
\end{tabular}
\caption[]{
{\sf
A comparison of predictions from {\tt ZFITTER} v.6.11 and {\tt TOPAZ0}
v.4.4 for muonic cross sections and forward-backward
asymmetries around the $Z$ peak.
First row is without initial-final state interference, second row with,
third row the relative effect of that interference in per mil
\cite{DESY99037,Jack:1999qh,Jack:1999ox}.
}}
\label{tab-acol10-th40}
\end{center}
\end{table}
%xxxxxxxxxxxxxxxxxxxxxxxxxxxxxxxxxxxxxxxxxxxxxxxxxxxxxxxxxxxxxxxxx 

%==========================================================================
\section{\bf EW and QED corrections at LEP and LC energies - a comparison
\label{ewboxes}
}
%--------------------------------------------------------------------------
%
As already mentioned above,
the semi-analytical program {\tt ZFITTER} was originally developed for 
{\tt SM} predictions of cross sections and asymmetries at LEP~1 energies.
Observables like total cross sections and asymmetries 
can be calculated in an {\it effective Born description},
as is done in the {\tt ZFITTER} approach: 
EW and QCD corrections are described as 
effective couplings in {\it effective Born observables} which are 
convoluted with the photonic corrections as flux functions. Higher 
order QED effects can then partly be described by resumming finite 
soft and virtual corrections ({\it soft-photon exponentiation}).   
This was illustrated e.g. in 
\cite{Bardin:1989all,Christova:1999cc,DESY99070,Christova:1998tc,DESY99037,Jack:1999qh,Jack:1999ox}.

\begin{figure}[htb]
\hspace*{-2cm} 
\begin{flushleft}
%--- 
\begin{tabular}{ll}
\hspace*{-0.5cm} 
  \mbox{%
  \epsfig{file=%
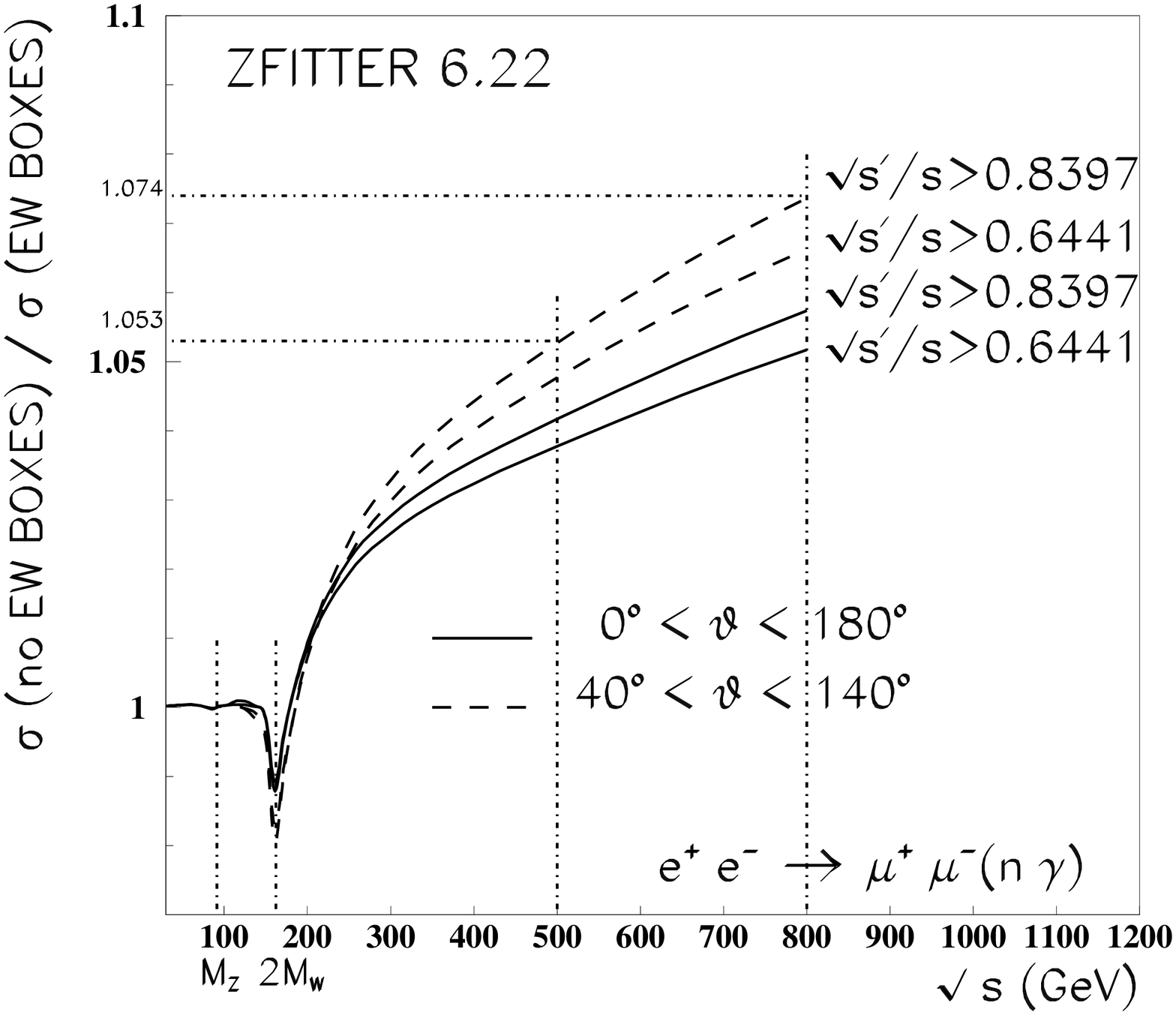
,width=8.cm   % this is the width of the figure (optional)
         }}%
&
  \mbox{%
  \epsfig{file=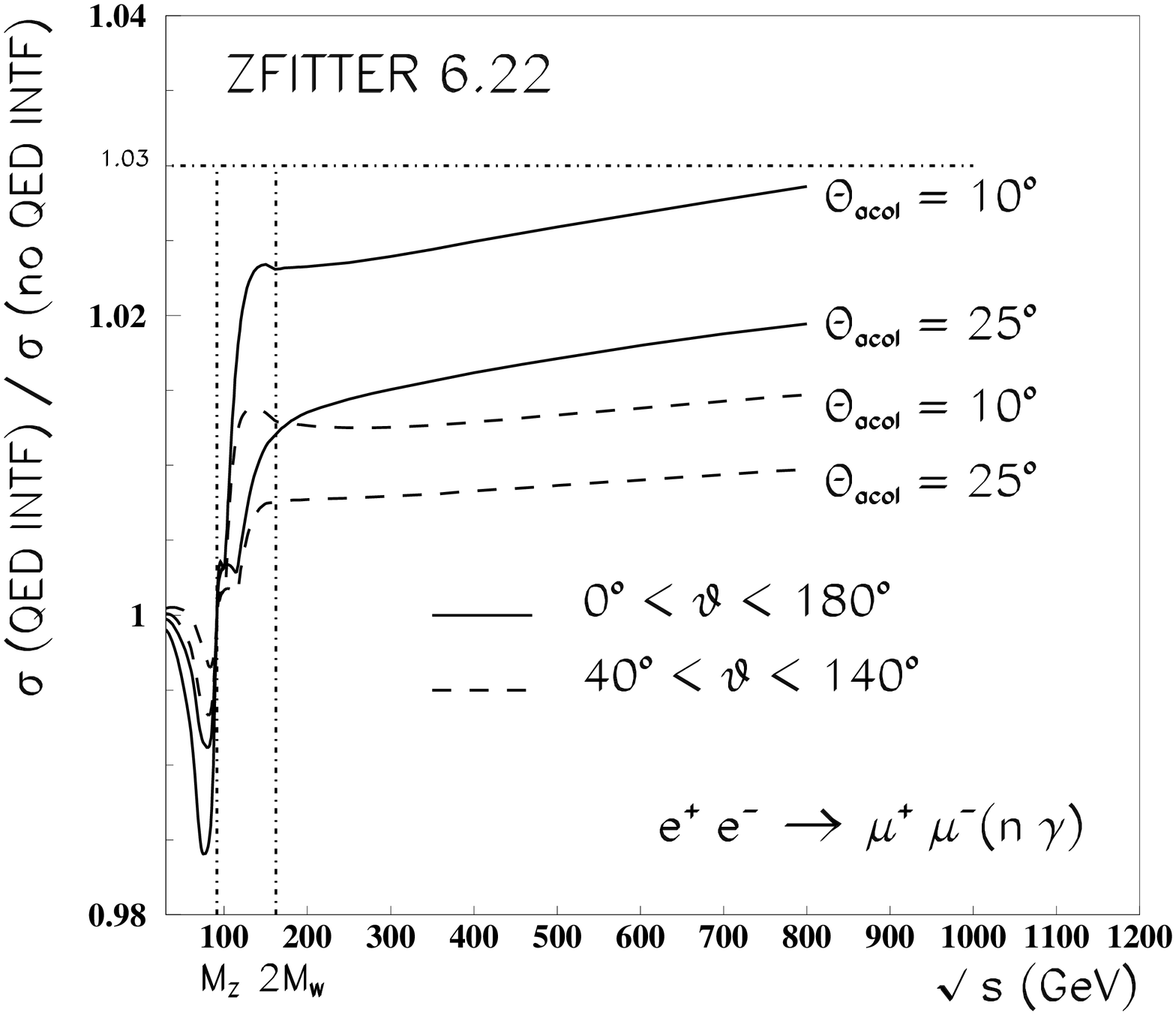,width=8.cm}}
\\
\end{tabular}
\caption[]
{\sf
a. Effect of EW box corrections at LEP and LC energies;
b. QED radiative corrections from initial-final state 
interference.
\label{sig_EW_QED}}
\end{flushleft}
\end{figure}

While at LEP 1 the EW and QCD corrections can in general be 
considered as small in comparison to the QED Bremsstrahlung,
this observation is not necessarily valid anymore at higher energies, 
where EW and QED corrections can grow to comparable magnitudes.
In order to underline this, we compared in Fig.~\ref{sig_EW_QED} 
%
%xxxxxxxxxxxxxxxxxxxxxxxxxxxxxxxxxxxxxxxxxxxxxxxxxxxxxxxxxxxxxxxxx
%xxxxxxxxxxxxxxxxxxxxxxxxxxxxxxxxxxxxxxxxxxxxxxxxxxxxxxxxxxxxxxxxx 
%
for muon-pair production cross sections $\sigma_T$ 
as an illustrative example the effect of virtual $ZZ$ and $WW$ box 
corrections as important EW corrections with corrections from 
the QED initial-final state interference.

In Fig.~\ref{sig_EW_QED}a, we switched off the $ZZ$ and $WW$ 
box corrections in order to visualize a positive effect.
Correspondingly, the QED interference was switched on and off 
in the right-hand plot (Fig.~\ref{sig_EW_QED}b). 
The net effect of these EW box corrections 
grows with increasing c.m.~energy roughly up to per cent 
level at LEP 2 energies, with the QED interference corrections
being slightly larger depending on the cut applied.
At LC energies, however, the EW contributions can even surpass
the QED interference contribution by roughly a factor of 2, while 
the effect from the QED interference approaches a more or less 
constant value of 2 to 3\%. 
For this comparison the {\tt ZFITTER} code, version v.6.22 
\cite{DESY99070}, was run `blindly' as it stands, i.e. without 
considering possible extra effects above the $t\bar{t}$ threshold 
due to the top quark mass. 
For an estimate of the EW situation at energies up to 1 TeV 
also consult for example \cite{Ciafaloni:1999sg}.

%==========================================================================
\section{\bf Photonic Corrections above the $Z$ resonance 
\label{qed_lc}   
}
\begin{figure}[htb]
\hspace*{-2cm}  
\begin{flushleft}
%--- 
\begin{tabular}{ll}
\hspace*{-0.5cm} 
  \mbox{%
  \epsfig{file=%
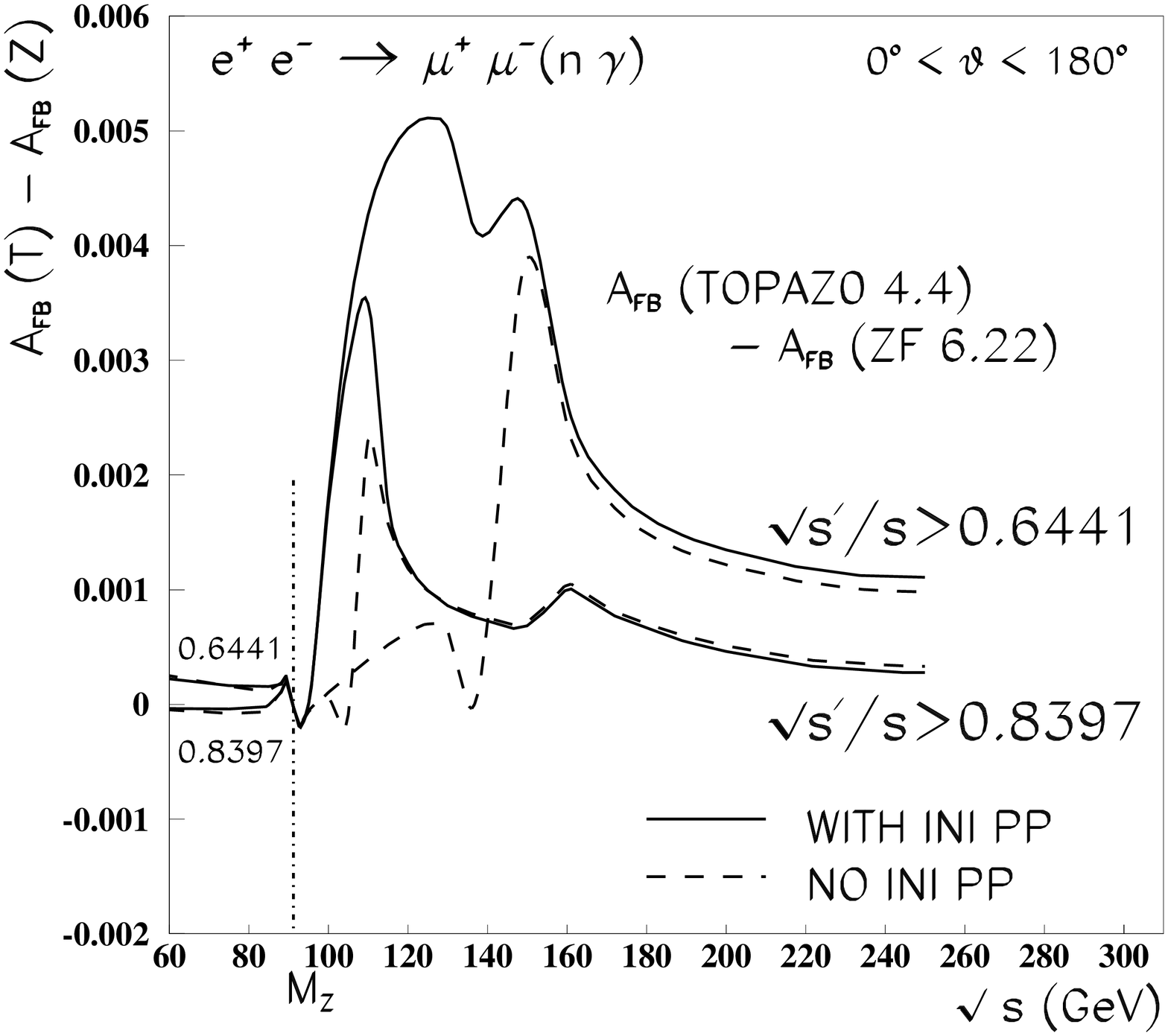,width=8.cm   % this is the width of the figure (optional)
         }}%
&
  \mbox{%
  \epsfig{file=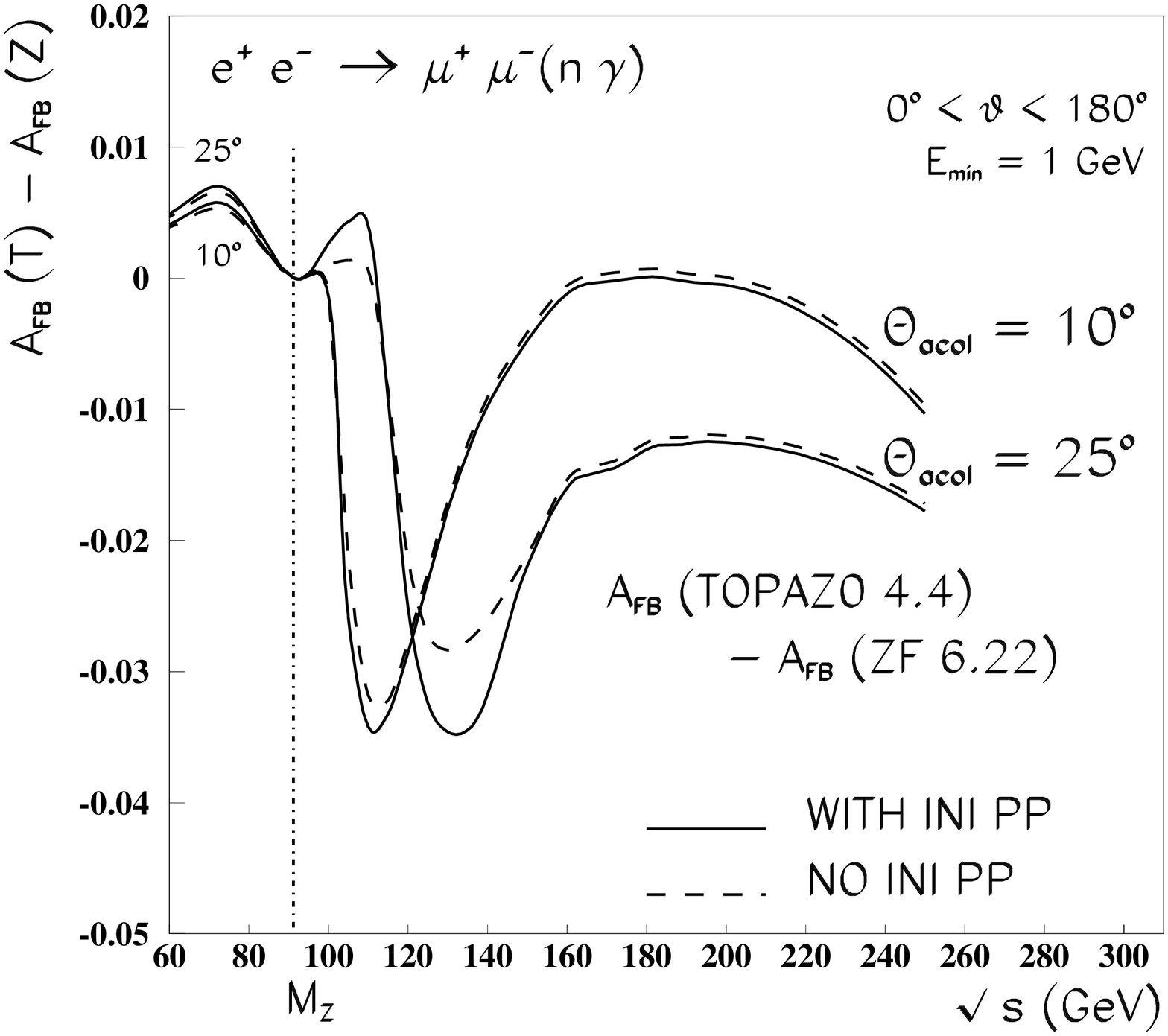,width=8.cm}}
\\
\end{tabular}
\caption[]
{\sf
Comparison of predictions from  {\tt ZFITTER} v.6.22 and 
{\tt TOPAZ0} v.4.4 for muon-pair forward-backward
asymmetries with a. an $s'$ cut or b. an acollinearity cut 
(Flag setting: {\tt ISPP}=0,1, {\tt FINR}=0; further: 
{\tt SIPP}={\tt S\_PR} \cite{DESY99070}). 
\label{afb-top-zf}}
\end{flushleft}
\end{figure}

%--------------------------------------------------------------------------
%
We want to focus now on the QED radiative corrections at 
higher energies. The status of the description of QED radiative 
corrections in the {\tt ZFITTER} code in comparison with other programs 
now at LEP 2 and higher energies below the $t\bar{t}$ threshold 
was examined in \cite{Christova:1998tc,Bardin:1999gt,SITGES99,Jack:1999qh,Jack:1999ox}: 
For the option with invariant mass cut the deviation of 
the codes {\tt ZFITTER} v.6.22 \cite{DESY99070} 
and {\tt TOPAZ0} v.4.4 \cite{Montagna:1998kp}
is typically not more than few per mil at LEP 2 energies
\footnote{For this, a sufficiently large invariant mass cut 
preventing the radiative return to the $Z$ boson resonance 
is applied.}
and is under control with respect to the experimentally 
demanded accuracy
\cite{Bardin:1999gt}.
For the acollinearity cut branch, 
the deviation of the codes increases to few per cent, 
even with stringent hard-photon cuts
\cite{SITGES99,Jack:1999qh,Jack:1999ox}. This is depicted in 
Fig.~\ref{afb-top-zf} for forward-backward asymmetries $A_{FB}$ 
of muon-pairs with different cuts 
(see also Fig.~16b. of \cite{DESY99037}).
%
%xxxxxxxxxxxxxxxxxxxxxxxxxxxxxxxxxxxxxxxxxxxxxxxxxxxxxxxxxxxxxxxxx 
%xxxxxxxxxxxxxxxxxxxxxxxxxxxxxxxxxxxxxxxxxxxxxxxxxxxxxxxxxxxxxxxxx 

\begin{figure}[htb]
\hspace*{-2cm}  
\begin{flushleft}
%--- 
\begin{tabular}{ll}
\hspace*{-0.5cm} 
  \mbox{%  
\epsfig{file=%
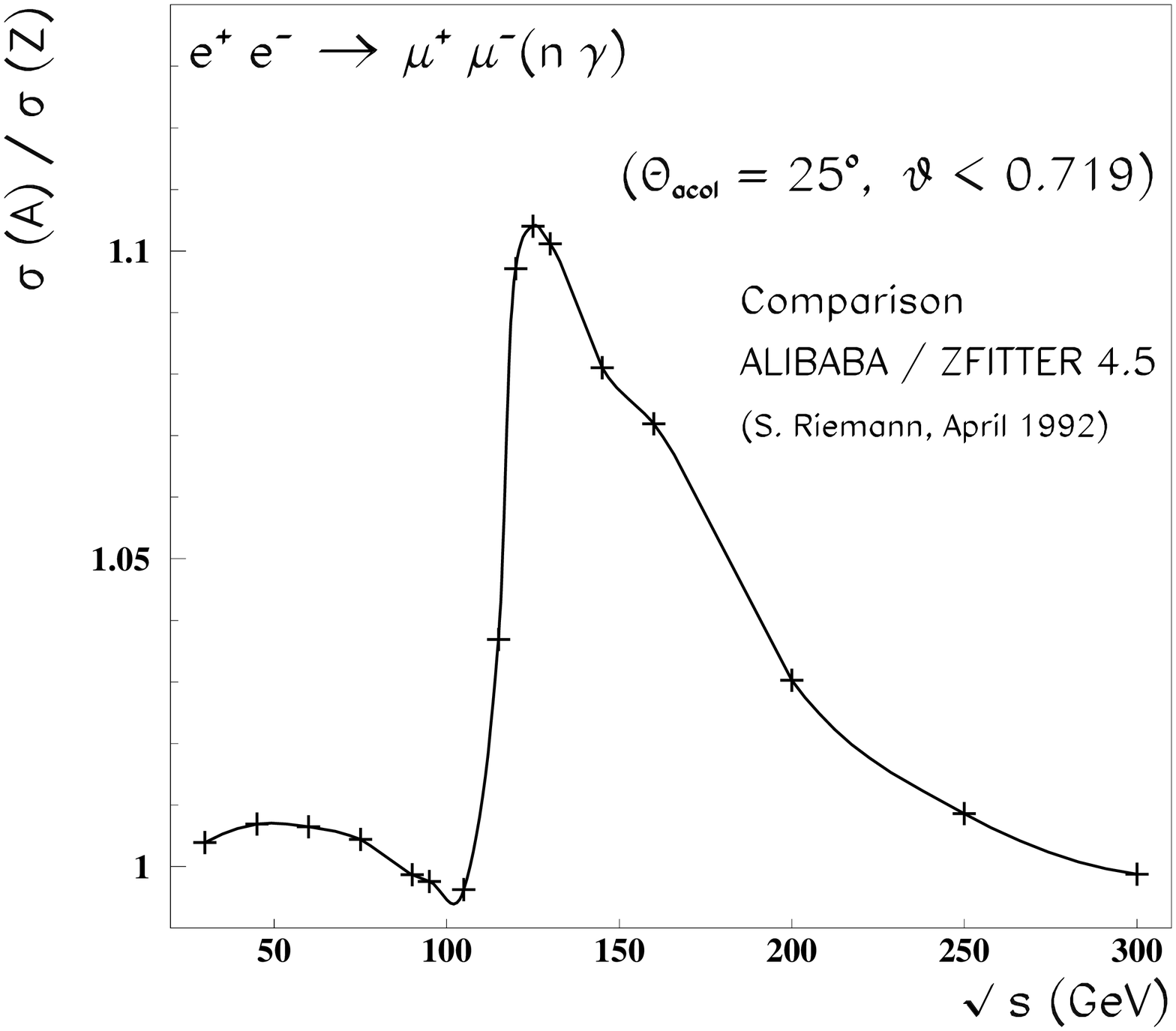,width=8.cm   % this is the width of the figure (optional)
         }}%
&
  \mbox{%
  \epsfig{file=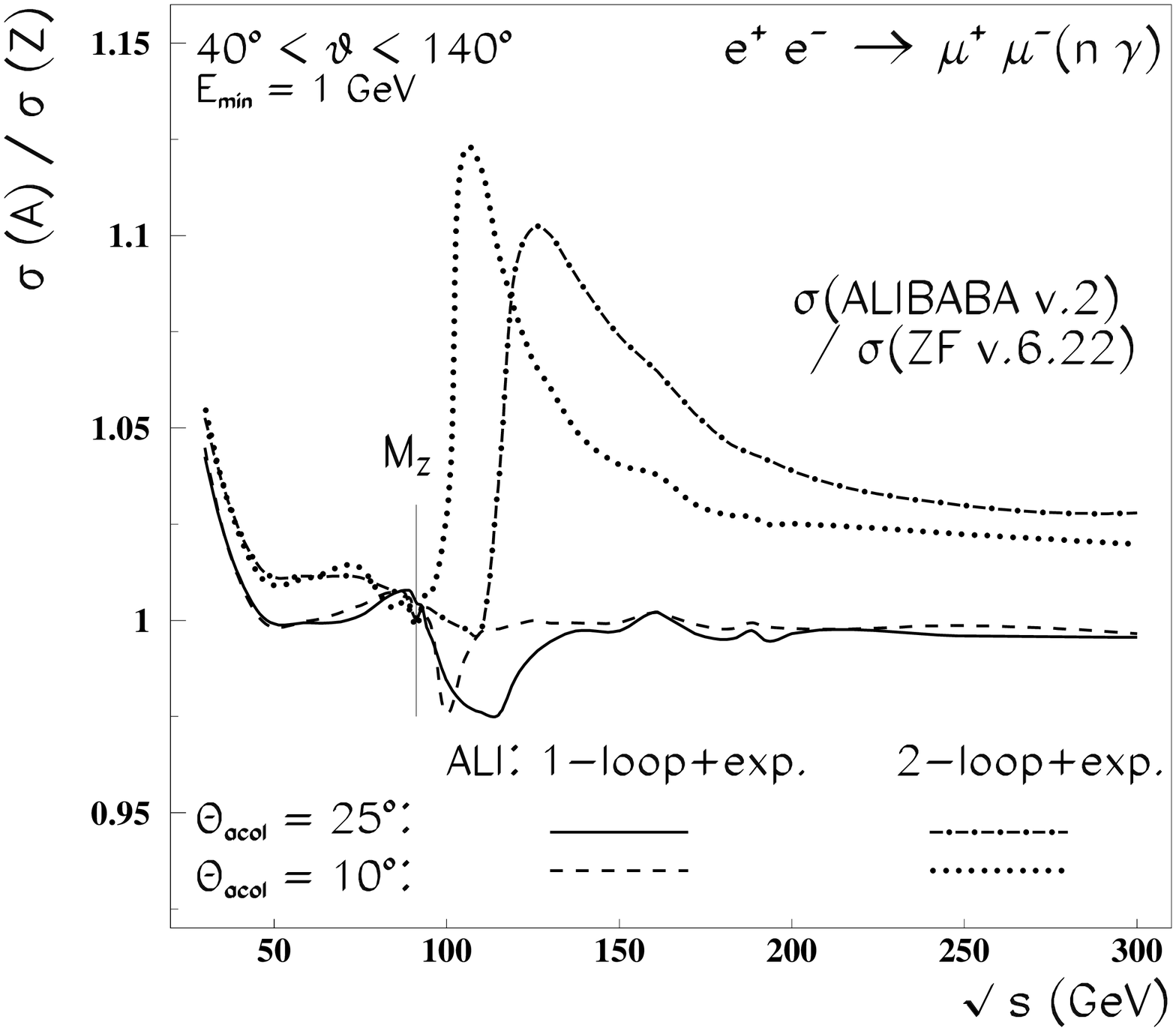,width=8.cm}}
\\
\end{tabular}
\caption[]
{\sf
Cross section ratios for muon-pair production with 
different acollinearity cuts from 30 to 800 GeV c.m.~energy;
Fig.a: Earlier comparison of {\tt ALIBABA} v.1 and {\tt ZFITTER} v.4.5 
\cite{Riemann:199200};
Fig.b: Updated comparison of codes: {\tt ALIBABA} v.2 and {\tt ZFITTER} v.6.22.
\label{ali_zf_codes}}
\end{flushleft}
\end{figure}

In Fig.~\ref{ali_zf_codes} we present an earlier analysis \cite{Riemann:199200} 
for the branch with acollinearity cut up to energies of 300 GeV for codes 
{\tt ALIBABA} v.1 \cite{Beenakker:1991mb} 
and {\tt ZFITTER} v.4.5 \cite{Bardin:1992jc},
together with an update of the comparison in \cite{DESY99037,SITGES99} 
for codes {\tt ALIBABA} v.2 \cite{Beenakker:1991mb} and 
{\tt ZFITTER} v.6.22~\cite{DESY99070}.
%
%xxxxxxxxxxxxxxxxxxxxxxxxxxxxxxxxxxxxxxxxxxxxxxxxxxxxxxxxxxxxxxxxx 
%xxxxxxxxxxxxxxxxxxxxxxxxxxxxxxxxxxxxxxxxxxxxxxxxxxxxxxxxxxxxxxxxx 
%
The per cent level discrepancy of the codes from LEP 2 
energies onwards has not changed much since then, although it seems now
clear that the large deviation is due to higher order QED corrections, 
contained in the code {\tt ALIBABA} v.1 and v.2, but not in {\tt ZFITTER} 
for the acollinearity cut branch \cite{Jack:1999qh,Jack:1999ox}.
The reason for these much larger deviations compared to the invariant 
mass cut is believed to be due to hard-photon effects from a radiative 
return to the $Z$ boson, not completely suppressed through an acollinearity 
cut and surviving at higher energies. This is still under investigation.

\begin{figure}[htb] 
\begin{flushleft}
%--- 
\begin{tabular}{ll}
\hspace*{1cm} 
  \mbox{%
  \epsfig{file=%
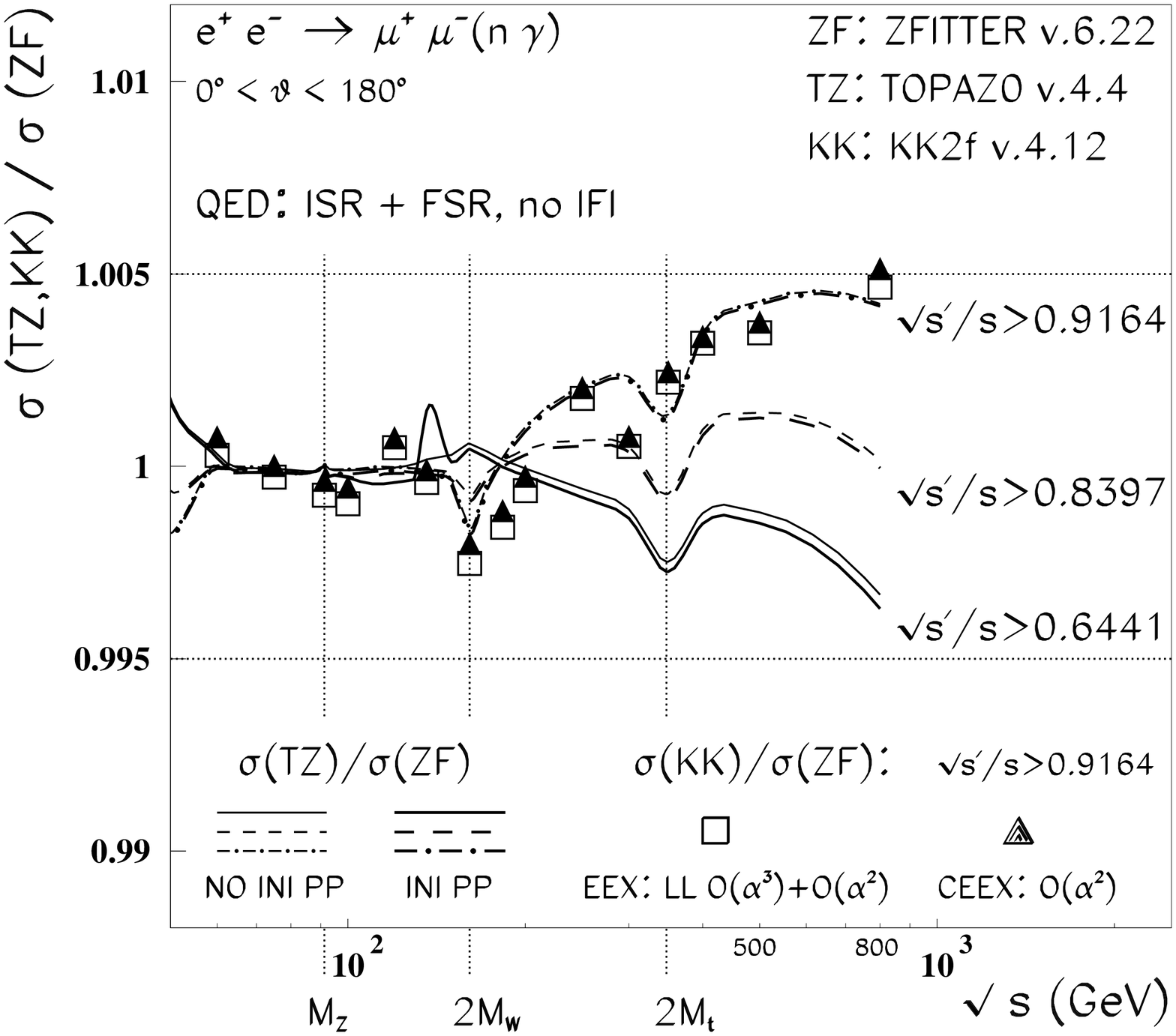,width=13.cm   % this is the width of the figure (optional)
         }}%
\end{tabular}
\vspace*{-1cm}
\caption[]
{\sf
Cross section ratios for muon-pair production with 
different $s'$ cuts
for codes {\tt ZFITTER} v.6.22, {\tt TOPAZ0} v.4.4, 
{\tt KK2f} v.4.12 (1999) from 60 to 800 GeV c.m.~energy;
without initial-final state interference
({\tt INI PP}: initial state pair production; {\tt LL}: leading logarithmic terms). 
\label{top_zf_kk_codes_a}}
\end{flushleft}
\end{figure}

\begin{figure}[htb] 
\begin{flushleft}
%--- 
\begin{tabular}{ll}
\hspace*{1cm} 
  \mbox{%
  \epsfig{file=%
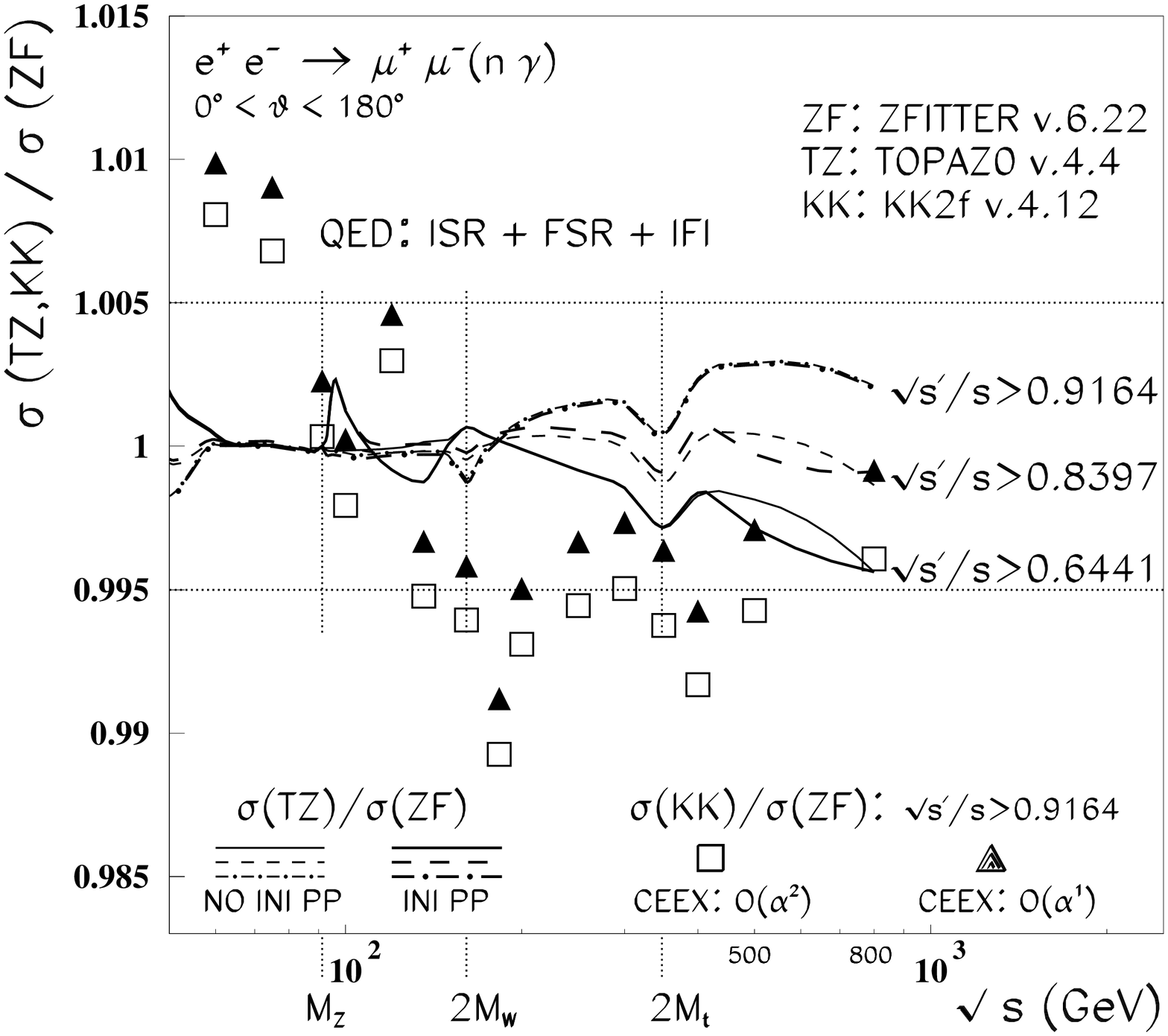,width=13.cm   % this is the width of the figure (optional)
         }}%
\end{tabular}
\vspace*{-1cm}
\caption[]
{\sf
Cross section ratios for muon-pair production with 
different $s'$ cuts for codes {\tt ZFITTER} v.6.22, {\tt TOPAZ0} v.4.4, 
{\tt KK2f} v.4.12 (1999) from 60 to 800 GeV c.m.~energy
with initial-final state interference
({\tt INI PP}: initial state pair production; {\tt LL}: leading logarithmic terms). 
\label{top_zf_kk_codes_b}}
\end{flushleft}
\end{figure}

We have now extended the comparison of total cross section predictions by 
codes {\tt ZFITTER} v.6.22, {\tt TOPAZ0} v.4.4, and {\tt KK2f} v.4.12 
\cite{Jadach:1999kkkz} up to typical LC energies of 500 to 800 GeV 
for the invariant mass cut option. This is shown in Figures  
\ref{top_zf_kk_codes_a} and \ref{top_zf_kk_codes_b}.
%
%xxxxxxxxxxxxxxxxxxxxxxxxxxxxxxxxxxxxxxxxxxxxxxxxxxxxxxxxxxxxxxxxx 
%xxxxxxxxxxxxxxxxxxxxxxxxxxxxxxxxxxxxxxxxxxxxxxxxxxxxxxxxxxxxxxxxx 
%
%xxxxxxxxxxxxxxxxxxxxxxxxxxxxxxxxxxxxxxxxxxxxxxxxxxxxxxxxxxxxxxxxx 
%xxxxxxxxxxxxxxxxxxxxxxxxxxxxxxxxxxxxxxxxxxxxxxxxxxxxxxxxxxxxxxxxx 

The general result of our analysis is that the deviation of the cross 
section predictions by the three codes is not more than 5 per mil 
for the complete energy range for the {\tt TOPAZ0}--{\tt ZFITTER} 
comparison. This observation also holds for the case of 
initial state QED Bremsstrahlung (ISR) alone 
when comparing code {\tt KK2f} with {\tt ZFITTER}, 
applying sufficiently strong invariant mass cuts and taking 
into account different higher order corrections 
(Fig.~\ref{top_zf_kk_codes_a}). Including QED initial-final 
state interference (IFI), our comparisons with {\tt KK2f} 
delivered a maximal deviation of roughly 1 \%
(Fig.~\ref{top_zf_kk_codes_b}).

In detail, this meant:
The numerical precision of {\tt TOPAZ0} and {\tt ZFITTER}
was better than $10^{-5}$ everywhere, while the accuracy  
of the Monte Carlo (MC) event generator {\tt KK2f} was necessarily
restricted due to limited CPU time: Calculating ISR with an accuracy of 
at least $10^{-3}$ required samples of 100000 events for each energy point. 
When including the (resummed) IFI, we had to use smaller samples of 30000   
events, resulting in a precision of only roughly $2\times 10^{-3}$. 
For ISR only, the typical CPU time per MC data point on e.g. 
an HP-UX 9000 workstation was about 25 minutes, increasing to 
roughly 100 minutes if IFI is added for the event samples stated 
above. In comparison, {\tt TOPAZ0} calculated one cross section value 
in a few minutes, while {\tt ZFITTER} with its semi-analytical approach 
calculated all 32 cross section values for one cut in a few seconds.
On the other hand, when interested in more complex setups, i.e. 
calculating multi-differential observables, using a wider variety of
cuts, or including extra higher order effects to the 
initial-final state interference, which {\tt ZFITTER} cannot or only 
partly provide, the numerical programs {\tt TOPAZ0}, or respectively 
{\tt KK2f}, clearly have their advantages.

We compared the effect of ISR solely (Fig.~\ref{top_zf_kk_codes_a}) or of 
ISR together with IFI (Fig.~\ref{top_zf_kk_codes_b}) for three different 
cut values, $\sqrt{s'/s} > 0.6441$, $0.8397$, and $0.9164$,
in the case of the {\tt TOPAZ0}--{\tt ZFITTER} comparison,  
and $\sqrt{s'/s} > 0.9164$ when comparing with {\tt KK2f}
\footnote{The cut values correspond approximately to a relatively strong 
cut on the maximal final state leptons' acollinearity angle of  
$25^\circ$, $10^\circ$, and $5^\circ$ respectively. 
}
with $s'$ defined here as the invariant mass squared of the $\gamma$ 
or $Z$ propagator after ISR, which is equal to the final state invariant 
mass squared including the emitted final state photons. For this, 
final state radition (FSR) was treated in form of a global correction 
factor. Alternatively, cutting on the final state invariant mass 
squared, $m^2_{f\bar{f}}$, after FSR, though it slightly worsened 
the good agreement at LEP 1 energies \cite{Bardin:1999gt},
it did not change the overall agreement in Figures 
\ref{top_zf_kk_codes_a} and \ref{top_zf_kk_codes_b} substantially.
For a recent discussion on this issue, defining 
kinematical cuts with radiative corrections  
for the experimental and computational situation
-- e.g. when including mixed QED and QCD contributions from
photonic and gluonic emission in the case of hadronic final states --
please consult \cite{Bardin:1999gt,Holt:LEP2EWWG}. 
In particular, at LEP 2 energies the predictions of the codes 
lie well inside the estimated experimental accuracies of  
e.g. $\Delta\sigma_{\mu\mu}\approx 1.2\%$, 
$\Delta\sigma_{had}\approx 0.5\%$ for sufficiently 
strong cuts \cite{Bardin:1999gt,Holt:LEP2EWWG}. 

Except where otherwise stated, we used the default settings of the 
programs, thus taking into account the $O(\alpha^2)$ photonic initial
state corrections with the leading logarithmic $O(\alpha^3)$ 
corrections together with the exact $O(\alpha)$ IFI contribution.
In {\tt ZFITTER} and {\tt TOPAZ0}, the $O(\alpha^2)$ corrections 
are complete. All three programs have installed resummed higher order 
corrections to ISR, exponentiating the finite soft and virtual 
photonic corrections.
\footnote{
The Yennie-Frautschi-Suura (EEX) prescription was used, resumming 
soft and virtual photonic contributions to all orders,
correctly removing all infrared divergences \cite{Jadach:1999kkkz}.
}
In contrast to codes {\tt ZFITTER} and {\tt TOPAZ0},
{\tt KK2f} also possesses a procedure to exponentiate 
IFI corrections with its newly implemented
{\it coherent exclusive exponentiation (CEEX)} 
\cite{Jadach:1998jb,Jadach:1999gz,Jadach:1999ob,Jadach:1999kkkz}.
\footnote{
The initial-final state interference contribution in {\tt KK2f} 
is only available with the {\it CEEX} option; for {\it EEX}
it is neglected.
}

To be more precise, {\it CEEX}  
does not include $O(\alpha^3)$ contributions up to now, 
while the {\it EEX} option in {\tt KK2f} does not contain 
the second order, subleading $O(\alpha^2 L)$ corrections, 
so both options are complementary to each other when interested in 
estimating these higher order effects for the initial state
Bremsstrahlung with {\tt KK2f}. 
In both cases {\tt KK2f} lacks the NNLL $O(\alpha^2 L^0)$ terms which, 
however, are estimated to be of the order $10^{-5}$ and so do not 
play a visible role in this comparison \cite{Jadach:1999kkkz}.

In Fig.~\ref{top_zf_kk_codes_a}, we compared the predictions by  
{\tt KK2f} with those by {\tt ZFITTER} for ISR alone, first for 
the {\it EEX} option with the LL $O(\alpha^3 L^3)$ and $O(\alpha^2 L^2)$
corrections ({\tt ZFITTER} flag values: {\tt FOT2} = 3, 5). Then we 
used the {\it CEEX} option for {\tt KK2f} and compared the $O(\alpha^2)$ results 
({\tt ZFITTER}: {\tt FOT2} = 2). The cross section ratios and the 
maximally 5 per mil deviation of the codes do not change considerably 
with values calculated with an uncertainty of $0.4\times 10^{-3}$.
at the $Z$ peak, and $1\times 10^{-3}$ overall.

In Fig.~\ref{top_zf_kk_codes_b} we give the cross section 
ratios, now with the IFI contribution, for {\it CEEX}
$O(\alpha^1)$ and {\it CEEX} $O(\alpha^2)$. There is roughly 
a 2 per mil shift of the central values -- always having in mind 
calculational uncertainties of $2\times 10^{-3}$ -- 
when going to the $O(\alpha^2)$ calculation, but they stay inside 
the overall $\pm 1$\% 
margin.

Other higher order corrections due to initial state 
pair creation, implemented in codes {\tt ZFITTER} and {\tt TOPAZ0},
only had minor effects on the cross section ratios 
at higher energies (see Fig.~\ref{top_zf_kk_codes_a} and 
\ref{top_zf_kk_codes_b}). It must be emphasized again that 
all three programs were run as they are, i.e. without 
considering perhaps necessary, later updates of the codes for  
electroweak corrections above the $t\bar{t}$ threshold.

Another important, in {\tt ZFITTER} recently updated contribution 
are initial state pair corrections if the Bremsstrahlung photon 
dissociates into a light fermion pair \cite{Bardin:LEP2MCW}: 
{\tt TOPAZ0} and {\tt ZFITTER} versions contain the $O(\alpha^2)$ 
leptonic and hadronic initial state pairs and a realization for simultaneous 
exponentiation of the photonic and pair radiators \cite{BKJho}. 
According to \cite{Jadach:1999kkkz}, initial state pair corrections 
are not included in the {\tt KK2f} code. Since {\tt ZFITTER} v.6.20, 
also the exact $O(\alpha^3)$ and the LL $O(\alpha^4)$ initial state pair 
corrections can be taken into account through convolution of the 
photonic and pair flux functions \cite{Arbuzov:1999}. The effect of the
pair corrections is e.g. with strong cuts roughly 2.5 per mil at the $Z$
peak, slightly decreases to approximately 2 per mil at LEP 2 energies,
and is not more than roughly 1 per mil at 500 to 800 GeV c.m. energy. 
In Fig.~\ref{top_zf_kk_codes_a} and \ref{top_zf_kk_codes_b}, switching on 
the pair corrections for different cuts, does not change the level of 
agreement between {\tt ZFITTER} v.6.22 and {\tt TOPAZ0} v.4.4
substantially. One interesting feature 
at lower energies between roughly 100 and 150 GeV -- just where the $Z$
radiative return is not prevented anymore by the applied cuts -- is 
the fact that the several per mil deviation of the two codes there
disappears when the pair corrections are switched off. From 
Fig.~\ref{top_zf_kk_codes_a} and \ref{top_zf_kk_codes_b} it can 
also be seen that such deviations can also be 
prevented by a sufficiently large $s'$ cut of e.g. $s'/s > 0.9$ 
if initial state pair corrections shall be included.

The inclusion of 4-fermion final states in this context, e.g. 
from final state pair creation, with their rather large, per cent 
level corrections at LEP 2 and higher energies \cite{Passarino:QFTHEP99}
naturally constitutes one of the next tasks which have to be approached
for an update of the codes for the LC.
Especially, the definition of background and signal diagrams in the 
hadronic case together with kinematical cuts -- experimentally 
and theoretically -- will be one of the major obstacles to
overcome \cite{Passarino:QFTHEP99}.
For a general summary of the present status of different available
codes at higher energies on 2-fermion, 4-fermion, $WW$ etc. physics 
see also \cite{Phillips:1999ox}.

%==========================================================================
\section{\bf ZFITTER above the $t\bar{t}$ threshold - a brief note 
\label{ttbar}
}
%--------------------------------------------------------------------------

The {\tt ZFITTER} code contains  
the complete one-loop virtual EW corrections to the 
$(\gamma,Z) f\bar{f}$ vertex from v.5.12 onwards
\cite{Akhundov:1986fct,Beenakker:1988pv,Bernabeu:1991ws}. 
While at LEP 1, the off-resonant $WW$ box corrections 
and the $\gamma b\bar{b}$ vertex corrections are negligible   
compared to the $Z b\bar{b}$ vertex, they become more
and more important with increasing c.m.~energy, especially 
at LEP 2 in the case of the $WW$ and $ZZ$ box corrections, 
leading to maximal effects between 2 and 4\%,
taking into account the $s$ dependency of the vertices 
and the angular dependency of the box corrections  
\cite{Christova:1998tc,LKreptoLEP98}. 
At higher energies, all virtual corrections will start to become
equally relevant introducing large gauge cancellations.
Corrections due to logarithmic and double
logarithmic ``Sudakov-type'' contributions 
from collinear and soft gauge boson exchange 
could lead to measurable 1\% 
or larger effects at a LC for $\sigma_T(b\bar{b})$ and 
$R_b$ at 500 GeV or higher, but only to per mil level
modifications for different ${b\bar{b}}$ asymmetries 
\cite{Beccaria:1999xd}.

The possibility of top quark pair production at a LC then 
was naturally one of the key fields of interest and 
discussions at this workshop \cite{Teubner:1999ox}. 
For the {\tt ZFITTER} code, this is one of the still missing, but 
urgently needed branches in order to make quick and reliable
estimates for $t\bar{t}$ cross sections with radiative 
corrections, at least for perturbative predictions sufficiently
above the $t\bar{t}$ threshold. A crucial role, of course, 
also plays the correct inclusion of mass effects at the 
available energies. While massive {\tt SM} and 
{\tt MSSM} calculations to $e^+e^-\to t\bar{t}$
with virtual and real QED \cite{Akhundov:1991a},
EW \cite{Beenakker:1991kh},
and QCD \cite{Ravindran:1998wvn} corrections
are already available,
work still has to be done concerning a description
in the context of EW form factors or the inclusion of 
hard QED and QCD corrections 
for the massive case \cite{BKNRL:1999??}. 

%==========================================================================
\section{\bf Conclusions
\label{sec_sum}
}
%--------------------------------------------------------------------------
%
Except for a possible and quite useful quick later run at a LC in the 
Giga-Z mode, the era of high precision measurements at the $Z$
boson resonance appears to be coming to an end. These high precision 
measurements have not only impressively confirmed the {\tt SM} predictions 
at the up-to-now available energies, but also led to many constraints on 
non-{\tt SM} physics, ranging from narrowing estimates on the mass of 
a light Higgs boson in the {\tt SM} or {\tt MSSM} to lower limits on 
new interaction scales. This is irrevocably also connected to the 
success of numerical codes for fermion-pair production like 
{\tt ZFITTER}, {\tt TOPAZ0}, {\tt KORALZ/KK2f}, and others, 
predicting experimentally measured quantities like cross sections and 
asymmetries with steadily increasing theoretical precision in order
to extract the interesting physics information from experimental data.

While the codes appear in `good shape' in the $Z$ boson resonance region 
with predictions at the per mil level or better 
\cite{Christova:1998tc,Bardin:1999gt,DESY99037,SITGES99,Jack:1999qh,Jack:1999ox}, 
still a lot has to be done for the higher energy domain at a $LC$ 
ranging e.g. from roughly 350 to 800 GeV, 
%for the {\tt TESLA} project \cite{TESLA}, 
if one takes into  
account the expected experimental precisions with high luminosities, 
polarized beams, and improved experimental analysis techniques.
Having this in mind, we have listed below some examples what updates 
we expect to become necessary for the semi-analytical program {\tt ZFITTER}
used at LC energies:
\begin{itemize}

\item For $A_{FB}$: LLA $O(\alpha^2)$ corrections from initial state 
      pair production;

\item Exponentiation of the initial-final state QED interference and 
      reexamination of the initial state exponentiation for diffferent cuts;

\item A {\tt ZFITTER}--{\tt GENTLE} \cite{Bardin:1996zz} merger: 
      final state pair corrections and other $4f$ contributions 
      (e.g. neutral current processes NC08, NC32);

\item Further comparisons between 
      {\tt TOPAZ0}--{\tt ZFITTER}--{\tt KK2f} etc.,
      especially for LEP 2, and then LC precisions;

\item $t\bar{t}$ production with real and virtual radiative corrections 
      including final state masses;

\item Further options like inclusion of beamstrahlung effects etc.
      \\ 
      \vdots 

\end{itemize}

With the combined effort of the different programming groups and constant 
interaction with the experimental community, this seems to be a tedious, 
but solvable `request list' of tasks with the rewarding promise
of delivering e.g. stringent mass bounds on a light Higgs boson, 
with a possible distinction between the {\tt SM} and {\tt MSSM} case,  
or giving hints for the mentioned, other extensions to the {\tt SM} 
-- or for even more `exotic, unasked-for' new physics.

%=========================================================================
\subsection*{Acknowledgments}
%--------------------------------------------------------------------------
We would like to thank D.~Bardin, L.~Kalinovskaya, S.~Jadach, and Z.~Was  
for helpful discussions.

%=========================================================================
\section*{\bf References}
%--------------------------------------------------------------------------

\end{document}